\documentclass[12pt]{article}
\usepackage[dvips]{color}
\usepackage{graphicx}
\usepackage{amssymb,amsmath,graphics,times}

\newcommand{\NaTimeConstant}{\tau_{ _{\rm Na}}}
\newcommand{\KTimeConstant}{\tau_{ _{\rm K}}}
\newcommand{\GTimeConstant}{\tau_{ _{\rm G}}}

\newcommand{\Kabattery}{E_{\rm K}}
\newcommand{\Naabattery}{E_{\rm Na}}
\newcommand{\Gabattery}{E_{\rm G}}

\newcommand{\Kconductance}{g_{ _{\rm K}}}
\newcommand{\Naconductance}{g_{ _{\rm Na}}}
\newcommand{\Gconductance}{g_{ _{\rm G}}}

\newcommand{\Kbarconductance}{{\bar g}_{ _{\rm K}}}
\newcommand{\Nabarconductance}{{\bar g}_{ _{\rm Na}}}
\newcommand{\Gbarconductance}{{\bar g}_{ _{\rm G}}}

\newcommand{\Kbarresistance}{{\bar r}_{ _{\rm K}}}
\newcommand{\Nabarresistance}{{\bar r}_{ _{\rm Na}}}
\newcommand{\Gbarresistance}{{\bar r}_{ _{\rm G}}}

\newcommand{\Kzero}{Q_{\rm K}}
\newcommand{\Nazero}{Q_{\rm Na}}
\newcommand{\Gzero}{Q_{\rm G}}
\newcommand{\Qzero}{Q}

\newcommand{\Keta}{\eta_{ _{\rm K}}}
\newcommand{\Naeta}{\eta_{ _{\rm Na}}}
\newcommand{\Geta}{\eta_{ _{\rm G}}}

\newcommand{\Kphi}{\phi_{ _{\rm K}}}
\newcommand{\Naphi}{\phi_{ _{\rm Na}}}
\newcommand{\Gphi}{\phi_{ _{\rm G}}}
\newcommand{\Xphi}{\phi_{ _{\rm X}}}

\newcommand{\Kpsi}{\psi_{ _{\rm K}}}
\newcommand{\Napsi}{\psi_{ _{\rm Na}}}
\newcommand{\Gpsi}{\psi_{ _{\rm G}}}
\newcommand{\Xpsi}{\psi_{ _{\rm X}}}

\newcommand{\Kgconductance}{\gamma_{ _{\rm K}}}
\newcommand{\Nagconductance}{\gamma_{ _{\rm Na}}}
\newcommand{\Ggconductance}{\gamma_{ _{\rm G}}}

\newcommand{\Vvoltage}{V}
\newcommand{\Vvoltageprime}{{V}'}

\begin{document}
%\centerline{\Large \sffamily On Mathematical Basis for Neuron Modeling}
%\vskip .1in

%\centerline{\large \sffamily Conductance-Resistance Symmetric Model for Excitable Membranes}

%\centerline{\Large \sffamily Conductance-Resistance Symmetric Model for Neuron}

%\centerline{\Large \sffamily Is Mathematics the Origin of Neuron?}

\centerline{\Large \sffamily Is Neuron Made from Mathematics?}

%\centerline{\Large \sffamily Is Neuron Made from Mathematical Symmetry?}

%\centerline{\Large \sffamily Is Neuron Made from Mathematical Symmetries?}

%\centerline{\Large \sffamily Is Neuron Made from Simple Symmetries?}

\bigskip
\centerline{\sffamily Bo Deng\footnote{Department of Mathematics,
University of Nebraska-Lincoln, Lincoln, NE 68588. Email: {\tt
bdeng@math.unl.edu}}}

\bigskip
\noindent{\bf Abstract: This paper is to derive a mathematical model for neuron by imposing only a principle of symmetry that two modelers must come up with the same model when one is approaching the problem by modeling the conductances of ion channels and the other by the channels' resistances.}

\bigskip
Because of its complexities no one thought it possible to derive mathematical models of neuron by logic alone. This paper is to show that perhaps is the case. The place to start is to assume an ion current through nerve cell's membrane to be Ohmic like $I=g(V-E)=\frac{1}{r}(V-E)$ and to ask if a modeler can derive the same model regardless whether she prefers to model the conductance $g$ or to model the resistance $r$. Here, $V$ is the intracellular membrane voltage, $E$ the ion species's Nernst potential, and $I$ the ion species's cross-membrane current.

These two approaches are constrained only by the conductance-resistance reciprocal symmetry:
\[%\begin{equation}\label{eqCRrelation}
gr=1.
\]%\end{equation}
As functions of the time, the conductance and resistance must satisfy by the chain rule that $r\frac{dg}{dt}+g\frac{dr}{dt}=0$.
The simplest assumption we can make about this relation is to assume the separation of variables equals a constant
\[
\frac{1}{g}\frac{dg}{dt}=-\frac{1}{r}\frac{dr}{dt}\equiv a
\]
for some scalar $a$. For $a=0$, it leads to the linear Ohmic channel $g\equiv$ constant, to which neural ion channels do not belong (\cite{cole1949dynamic}). For $a\ne 0$, either $g(t)=g_0e^{at}$ or $r(t)=r_0e^{-at}$ grows in time without bound. But this is not consistant with what we know about neurons or any natural process since at clamped voltages both potassium and sodium channels' conductances saturate at finite values (\cite{hodgkin1952quantitative}). We then assume instead that the righthand side be a function of the conductance (or equivalently the resistance)
\[
\frac{1}{g}\frac{dg}{dt}=A(g),\text{equivalently}, \frac{1}{r}\frac{dr}{dt}=-A(g).
\]
We further ask if there are functions $A(g)$ so that regardless a modeler's preference the two models look the same? That is, if there are functions $A(g)$ so that
\[
\frac{1}{g}\frac{dg}{dt}=A(g),\text{and}, \frac{1}{r}\frac{dr}{dt}=-A(g)=A(r)?
\]
If true, it imposes the following condition
\begin{equation}\label{eqCRsymmetry}
-A(g)=A(r),\text{equivalently}, A(g)+A(r)=0.
\end{equation}
It turns out a simple nonzero solution to the equation is of the form
\begin{equation}\label{eqCRsymmetry1}
A(g)=\tau\frac{1}{\sqrt{g\gamma}}(\gamma-g)
\end{equation}
for a two-parameter family of functions with parameters $\tau,\gamma$. It is straightforward to check
\[
-A(g)=-\tau\frac{1}{\sqrt{g\gamma}}(\gamma-g)=\tau\frac{1}{\sqrt{r\rho}}(\rho-r)=A(r)
\]
by renaming the parameter $\gamma=1/\rho$. The functional form (\ref{eqCRsymmetry1}) is referred to satisfy the  {\em conductance-resistance symmetry}.

To see how far this line of reasoning can go, we first solve the conductance kinetic equation
\[
\frac{dg}{dt}=gA(g)=\tau\sqrt{\frac{g}{\gamma}}(\gamma-g)
\]
by some undergraduate textbook techniques for ordinary differential equations.
The solution is
\[
g(t)=\gamma\left[\frac{ke^{\tau t}-1}{ke^{\tau t}+1}\right]^2,\ \text{with}\  k=\frac{1+\sqrt{g_0/\gamma}}{1-\sqrt{g_0/\gamma}}
\]
with $g(0)=g_0$ being the initial value.
The solution is very illuminating. For $\tau\ne 0$, $g(t)$ converges to $\gamma$ as $t\to +\infty$. 
For $\tau>0$ and $0<g_0<\gamma$, $g(t)$ is always increasing. This seems to suggest that if a voltage is clamped at a given value, the ion channel's conductance must saturate toward the value $\gamma$. In addition, the rate at which the convergence takes place is of $e^{-\tau t}$, implying that $\tau$ is exactly the time constant for the conductance kinetics. Because $r(t)=1/g(t)$ is the solution to the resistance equation $dr/dt=rA(r)$ we have the symmetric form for the solution
\[
r(t)=\rho\left[\frac{ke^{\tau t}+1}{ke^{\tau t}-1}\right]^2,\ \text{with}\  k=\frac{\sqrt{r_0/\rho}+1}{\sqrt{r_0/\rho}-1}.
\]
To emphasize the role of parameter $\gamma$ and $\rho$ we denote the equations by
\begin{equation*}%\label{eqCRsymmetry1}
\frac{dg}{dt}=gA(g)=\tau\sqrt{\frac{g}{\gamma}}(\gamma-g):=B(g,\gamma,\tau),\ \text{and}\ \frac{dr}{dt}=rA(r)=\tau\sqrt{\frac{r}{\rho}}(\rho-r)=B(r,\rho,\tau)
\end{equation*}
and say they satisfy the {\em conductance-resistance kinetic symmetry} (CRKS).

Since $\gamma,\rho$ are the voltage-clamped maximal conductance and minimal resistance respectively, they are functions of the cross-membrane voltage $V$ satisfying $\gamma(V)\rho(V)=1$. Differentiating the identity in $V$ we obtain similarly by separating the variables
\[
\frac{1}{\gamma}\frac{d\gamma}{dV}=-\frac{1}{\rho}\frac{d\rho}{dV}.
\]
We assume also these voltage-dependent $\gamma$ and $\rho$ satisfy a similar conductance-resistance symmetry with respect to the cross-membrane voltage instead. Then,
\begin{equation}\label{eqCRsymmetry2}
\frac{1}{\gamma}\frac{d\gamma}{dV}=A(\gamma)\hbox{ and } \frac{1}{\rho}\frac{d\rho}{dV}=A(\rho)
\end{equation}
with $A(\gamma)+A(\rho)=0$ for some functional $A$.

Two types of channels are treated separately: voltage-activation ion channel and voltage-gating channel. For the first type, we assume that the CR symmetric $A(\gamma)$ has the same functional form as (\ref{eqCRsymmetry1}) with a positive $V$-rate parameter. Specifically we have
\[
\frac{1}{\gamma}\frac{d\gamma}{dV}=A(\gamma), \hbox{\ with\ }A(\gamma)=\eta\frac{1}{\sqrt{\bar g\gamma}}(\bar g-\gamma)
\]
where $\eta>0$ is the generalized `time constant' respect to the cross-membrane voltage $V$, and $\bar g$ is the maximal conductance as $V$ increases to infinity. Namely, for the ion channel, the channel conductance $\gamma$ increases with depolarization in increasing $V$ because $A(\gamma)>0$ for $\gamma<\bar g$. Similarly, $\gamma$ decreases with hyperpolarization in decreasing $V$. Again, $\gamma$ can be solved explicitly as
\[
\gamma(V)=\bar g\left[\frac{ke^{\eta V}-1}{ke^{\eta V}+1}\right]^2,\ \text{with}\  k=\frac{1+\sqrt{\gamma_0/\bar g}}{1-\sqrt{\gamma_0/\bar g}}
\]
with the $\gamma_0$ being the `initial' conductance when $V=0$, and the property that $\lim_{V\to\infty}\gamma(V)=\bar g$. Moreover, this solution holds at least for $V\ge 0$.

To extend the solution below $V=0$, we need to note a few facts about the equation
\[
\frac{d\gamma}{dV}=\eta\sqrt{\frac{\gamma}{\bar g}}(\bar g-\gamma)=B(\gamma,\bar g, \eta).
\]
First, it has the trivial solution $\gamma(V)\equiv 0$. Second, a solution is increasing (or non-decreasing) in $V$ if it is below $\bar g$ at some value of $V$. Most important of all, because the right hand is not differentiable at $\gamma=0$, the solution may not be unique when originated from $\gamma=0$. In fact, we can explicitly construct another solution which is zero for $V$ below some value, $\Qzero$, and strictly increasing above $\Qzero$. More specifically, we can re-parameterize and rewrite the solution above as
\[
\gamma(V)=\bar g\left[\frac{ke^{\eta V}-1}{ke^{\eta V}+1}\right]^2
=\bar g\left[\frac{e^{\eta (V-\Qzero)}-1}{e^{\eta (V-\Qzero)}+1}\right]^2
\]
with $\Qzero=-\ln k/\eta$. Then $\gamma(V)$ exists for $V\ge \Qzero$ and more importantly, $\gamma(\Qzero)=0$. Notice that this form can be further simplified as
\[
\gamma(V)=\bar g\tanh^2\left(\frac{\eta}{2}(V-\Qzero)\right).
\]
By further extending this solution below $\Qzero$ to be $\gamma(V)=0$ we obtain the solution we need
\begin{equation}\label{eqCRsymmetry3}
\gamma=\bar g\tanh^2\left(\frac{\eta}{2}(V-\Qzero)\right)H(V-\Qzero):=\bar g\phi(V,\eta,\Qzero)
\end{equation}
where $H(x)$ is the Heaviside function with $H(x)=0,\ x<0$ and $H(x)=1,\ x\ge 0$. Notice more importantly that the function $\phi(V,\eta,\Qzero)=\tanh^2\left(\frac{\eta}{2}(V-\Qzero)\right)H(V-\Qzero)$ whose range is $[0,1)$ can be interpreted as the probability of opening pours for the ion. For sodium and potassium ion channels, we have their corresponding limiting conductances as
\[
\Kgconductance=\Kbarconductance\Kphi(V,\Keta,\Kzero),\hbox{ and } \Nagconductance=\Nabarconductance\Naphi(V,\Naeta,\Nazero).
\]

The phenomenon of voltage-gating (\cite{armstrong1973currents}) occurs when a small pulse-like outward current is generated due to the release of charged molecules from the sodium channel pores in responding to some conformational changes of the pores to depolarizing voltage. Its effect is opposite to voltage-activated ion channels. That is, unlike ion channels, gating conductance decreases with depolarizing voltage and increases with hyperpolarization. Again, we assume the gating channel is Ohmic-like whose time-dependent conductance satisfies CRKS for which the voltage-dependent limiting conductance is also conductance-resistance symmetric satisfying (\ref{eqCRsymmetry2}) but with a negative $V$-rate constant. Specifically we have 
\[
\frac{1}{\gamma}\frac{d\gamma}{dV}=A(\gamma)=-\eta\frac{1}{\sqrt{\bar g\gamma}}(\bar g-\gamma).
\]
Since $A(\gamma)+A(\rho)=0$, it is straightforward to check 
\[
\frac{1}{\rho}\frac{d\rho}{dV}=A(\rho),
\]
showing in fact the conductance-resistance symmetry is satisfied. For conductance equation, we can derive or check by exactly the same arguments above that it has a solution $\Ggconductance=\Gbarconductance\Gphi(V,\Geta,\Gzero)$ with $\Gphi$ defined as follows
\[
\Gphi(V,\Geta,\Gzero)=H(\Gzero-V)\tanh^2\left(\frac{\Geta}{2}(\Gzero-V)\right),
\]
which is a decreasing function for $V\le \Gzero$ and zero for $V\ge \Gzero$.

%%%%%%%%%%%%%%%%%%%%%%%%%%%%%%%%%%%%%%%%%%%%%%%%%%%%%%%%%%%%%
%%%%%%%%%%%%%%%%%%%%%%%%%%%%%%%%%%%%%%%%%%%%%%%%%%%%%%%%%%%%%
%%%%%%%%%%%%%%%%%%%%%%%%%%%%%%%%%%%%%%%%%%%%%%%%%%%%%%%%%%%%%
\begin{figure}%[ht]

\centerline{
\parbox[l]{3in}{
\centerline
{\scalebox{.5}{\includegraphics{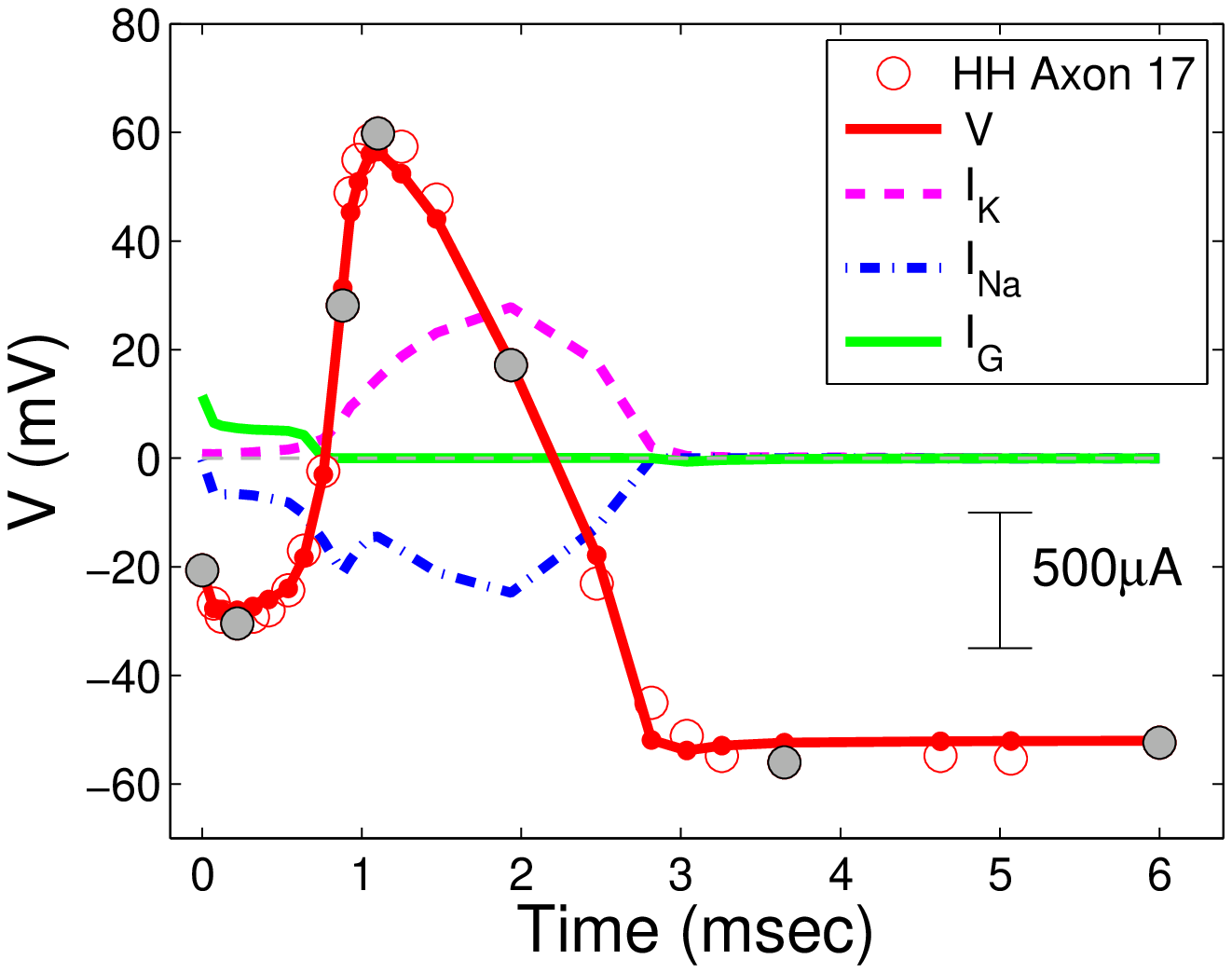}}}
}
\hskip 1cm
\parbox[l]{3in}{
\centerline
{\scalebox{.5}{\includegraphics{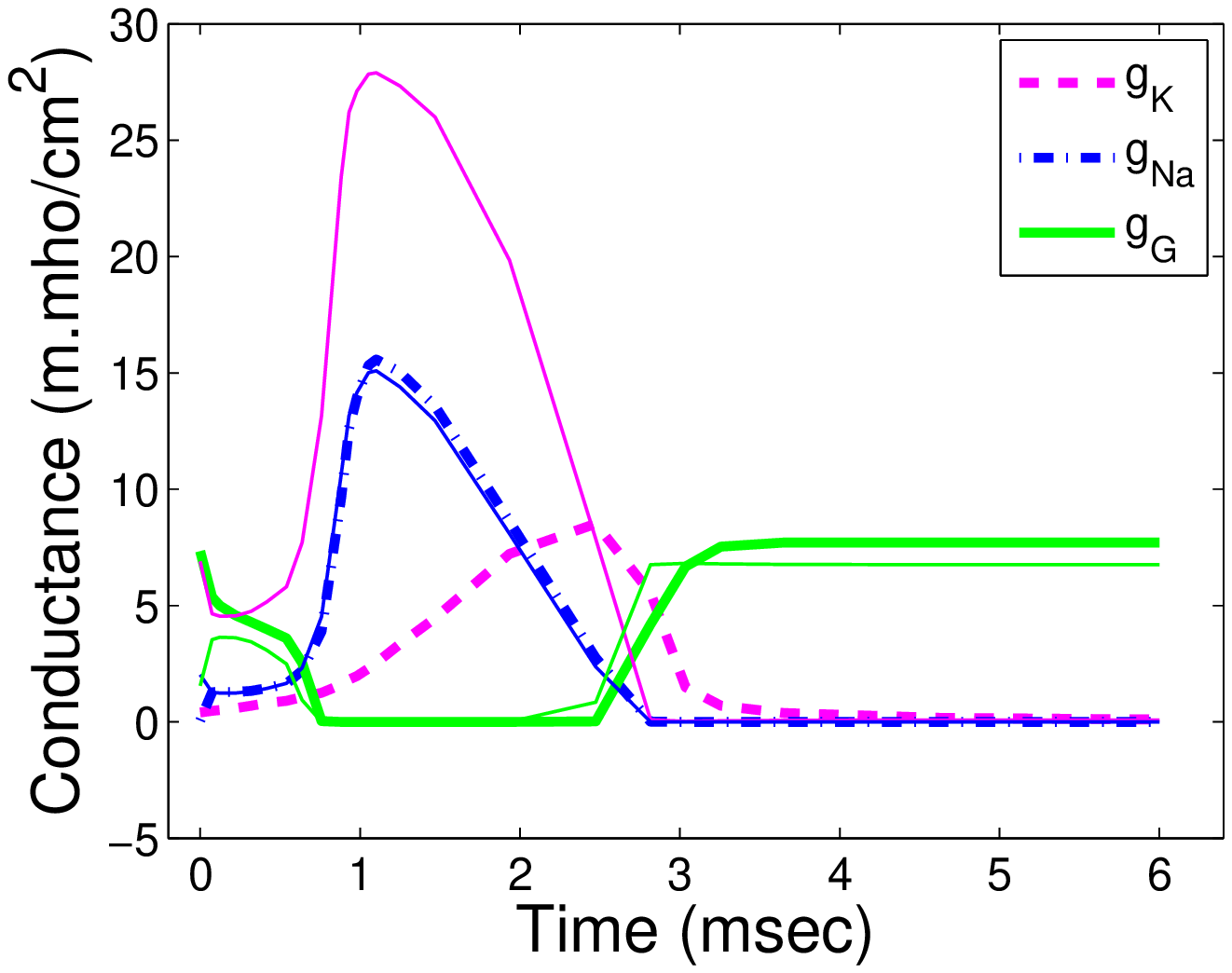}}}
}
}
\vskip .07in
\centerline{\ \ \ \ (a)\hskip 3.3in (b)}

\centerline{
\parbox[l]{3in}{
\centerline
{\scalebox{.5}{\includegraphics{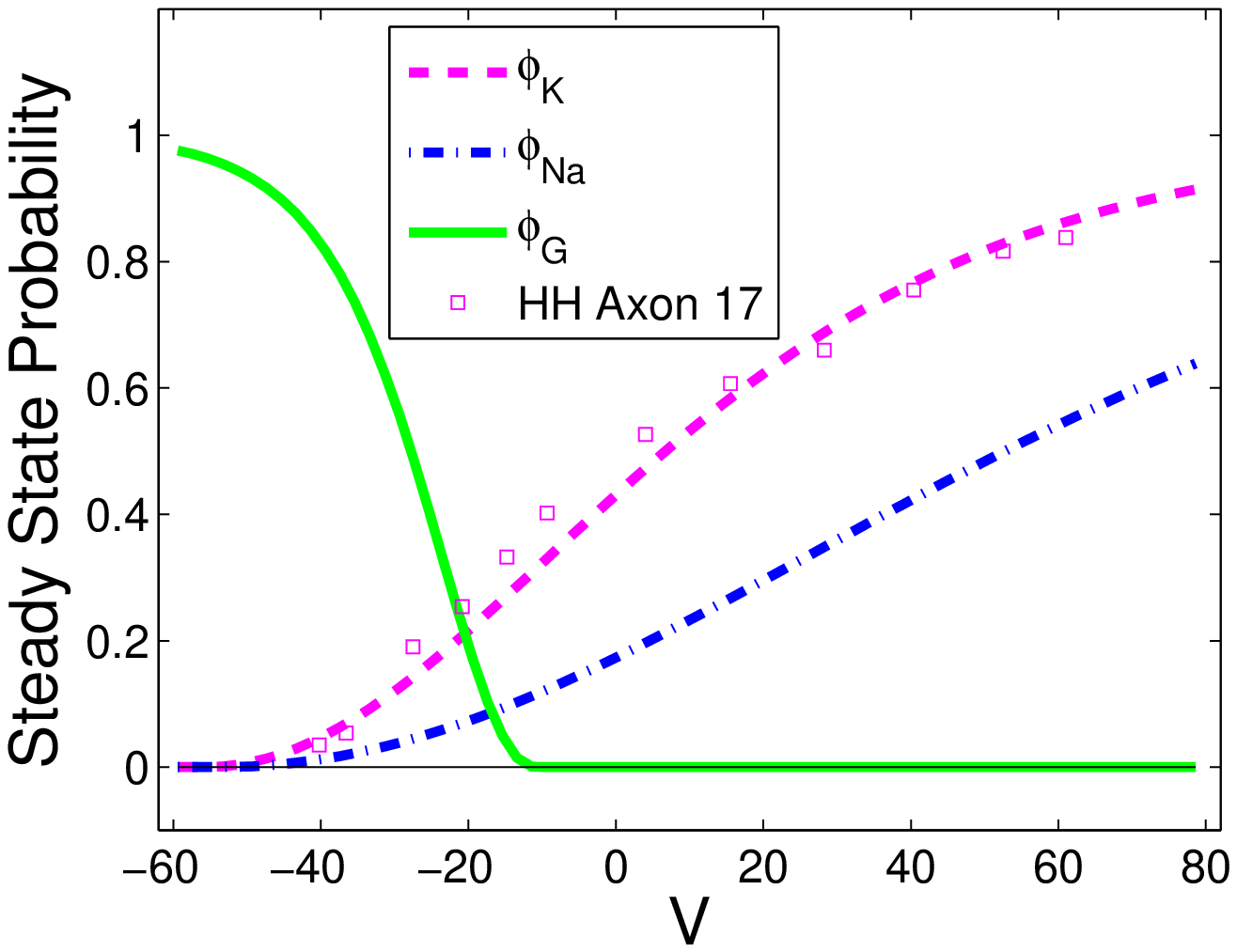}}}
}
\hskip 1cm
\parbox[l]{3in}{
\centerline
{\scalebox{.5}{\includegraphics{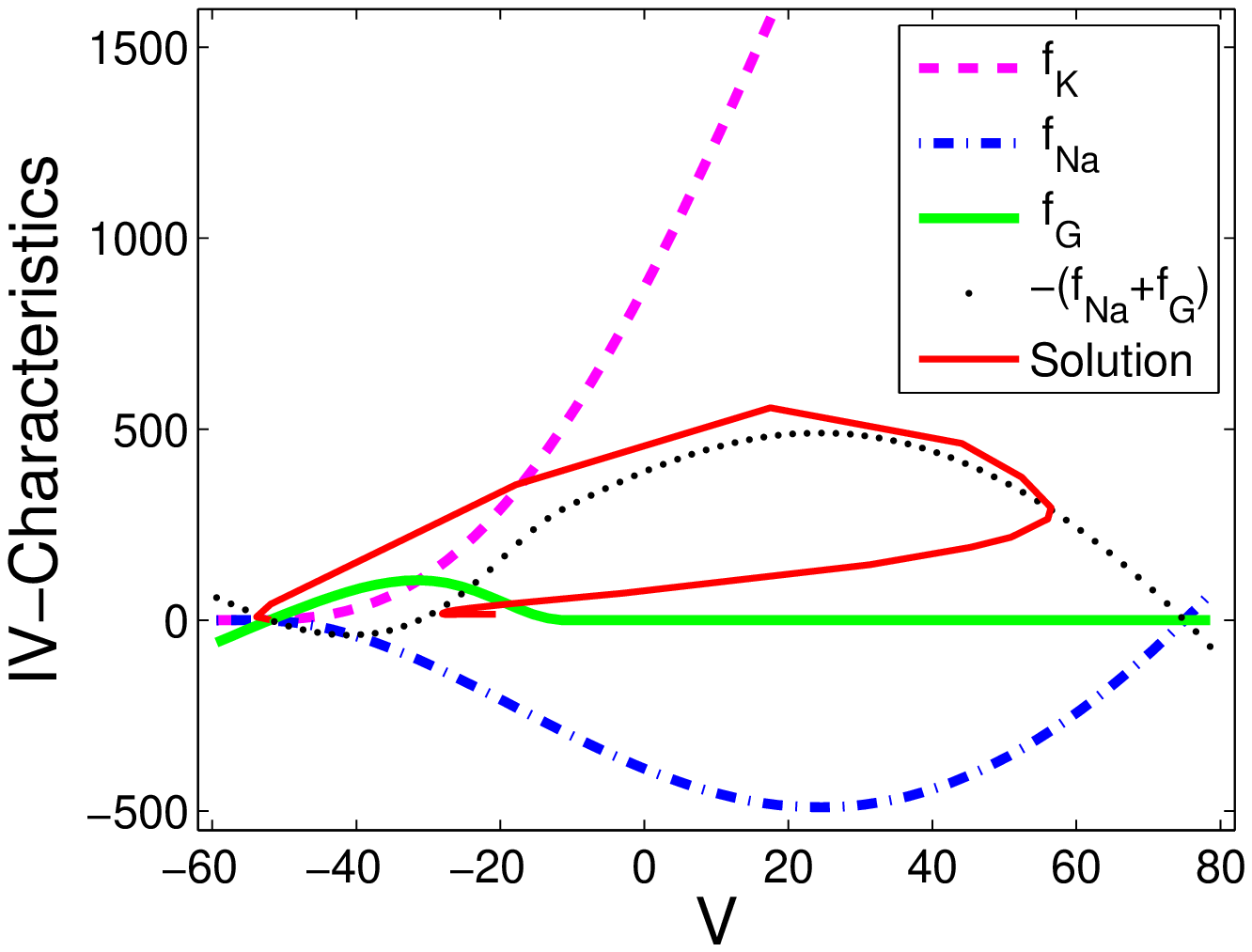}}}
}
}
\vskip .07in
\centerline{\ \ \ \ (c)\hskip 3.3in (d)}

\caption{(a) Best-fit parameter values for (\ref{eqDengConductanceModel}) are: $\Kabattery=-59.5$ mV, $\Kbarconductance=34$ m.mho/cm$^2$, $\Kzero=-55.6$ mV, $\Keta=0.02830$, $\Naabattery=75$ mV, $\Nabarconductance=29.92009$ m.mho/cm$^2$, $\Nazero=-53.31456$ mV, $\Naeta=0.01662$, $\Gabattery=-52.0000$ mV, $\Gbarconductance=8.12495$ m.mho/cm$^2$, $\Gzero=-11.08610$ mV, $\Geta=0.10566$, $C=1\mu{\rm F/cm}^2$, $\KTimeConstant=0.59167$/msec, $\NaTimeConstant=42.93673$/msec, and $\GTimeConstant=6.45857$/msec. Also, $\epsilon=\delta=10^{-4}$. The initial values are $V_0=-20.67$, a depolarized value from the resting potential, and $n_0=\Kphi(V_0)$, $m_0=\Naphi(V_0)$, $h_0=\Gphi(V_0)$. Filled disks are the data points used for the best fit. (b) The conductances $g_{ _X}$ as functions of the time for the action potential. Thin lines are the corresponding characteristic conductances, namely $\bar g_{ _X}\phi_{ _X}$. (c) The steady-state characteristic conductances $\phi_{ _X}$ are functions of $V$. In fact, parameters $\Kzero, \Keta$ are best-fitted to the experiment data of Fig.5 of \cite{hodgkin1952quantitative}, and the rest of the parameters of (a) are best-fitted to the experiment data of Fig.12 of \cite{hodgkin1952quantitative}. In particular, parameter values $\Kabattery, \Naabattery, \Kbarconductance$ are taken from \cite{hodgkin1952quantitative} and the rest are best-fitted by our gradient search algorithm. (d) $IV$-characteristics for ion currents, e.g., $I=f_{\rm K}(V)$ is given by $f_{\rm K}(V)=\Kbarconductance \Kphi(V)(V-\Kabattery)$ etc. Dotted line is the negative of the combined sodium and gating characteristics, $I=-(f_{\rm Na}(V)+f_{\rm G}(V))$. The resting membrane potential equilibrium is the intersection of this curve with the potassium characteristic curve $I=f_{\rm K}(V)$.}\label{figBestModelFit}
\end{figure}
%%%%%%%%%%%%%%%%%%%%%%%%%%%%%%%%%%%%%%%%%%%%%%%%%%%%%%%%%%%%%
%%%%%%%%%%%%%%%%%%%%%%%%%%%%%%%%%%%%%%%%%%%%%%%%%%%%%%%%%%%%%
%%%%%%%%%%%%%%%%%%%%%%%%%%%%%%%%%%%%%%%%%%%%%%%%%%%%%%%%%%%%%

Surprisingly the conductance models satisfying the kinetic symmetry and the activation-gating symmetry are unit free or scale-invariant. For example, for the potassium channel we can rescale $\Kconductance =\Kbarconductance n$ to simplify the kinetic equation $d\Kconductance/dt=B(\Kconductance,\Kgconductance,\KTimeConstant)$ to
\[
n'=\frac{dn}{dt}=B(n,\Kphi,\KTimeConstant).
\]
Similarly, for the sodium and gating currents the rescaling $\Naconductance =\Nabarconductance m$, $\Gconductance =\Gbarconductance h$ give
\[
m'=B(m,\Naphi,\NaTimeConstant),\quad h'=B(h,\Gphi,\GTimeConstant).
\]
As a result when we couple the conductance kinetics together with the voltage kinetics by the Kirchhoff current law we obtain the following possible model if we prefer to model the membrane channels by their conductances:
\begin{equation}\label{eqDengGRModel}
\qquad\qquad\qquad\left\{\!\!\!\!\begin{array}{ll} & C\Vvoltageprime=-[\Kbarconductance n (\Vvoltage-\Kabattery)+\Nabarconductance m (\Vvoltage-\Naabattery)+\Gbarconductance h (\Vvoltage-\Gabattery)]\\
& {n}' = \displaystyle \KTimeConstant\sqrt{n/{\Kphi}}(\Kphi-n)\\
& {m}' = \displaystyle \NaTimeConstant\sqrt{m/{\Naphi}}(\Naphi-m)\\
& {h}' = \displaystyle \GTimeConstant\sqrt{h/{\Gphi}}(\Gphi-h)\\
\end{array}\right.
\end{equation}
with $\Kphi,\Naphi,\Gphi$ as the voltage-dependent probabilities given above.
This model looks exactly the same if we choose to model the channels by their resistances with ${\bar r}_{ _{\rm X}}=1/{\bar g}_{ _{\rm X}}$, $\Xpsi=1/\Xphi$, $x=1/n$, $y=1/m$, $z=1/h$,
\begin{equation}\label{eqDengGRModel2}
\qquad\qquad\qquad\left\{\!\!\!\!\begin{array}{ll} & C\Vvoltageprime=-[(\Vvoltage-\Kabattery)/(\Kbarresistance x)+(\Vvoltage-\Naabattery)/(\Nabarresistance y)+(\Vvoltage-\Gabattery)/(\Gbarresistance z)]\\
& {x}' = \displaystyle \KTimeConstant\sqrt{x/{\Kpsi}}(\Kpsi-x)\\
& {y}' = \displaystyle \NaTimeConstant\sqrt{y/{\Napsi}}(\Napsi-y)\\
& {z}' = \displaystyle \GTimeConstant\sqrt{z/{\Gpsi}}(\Gpsi-z).\\
\end{array}\right.
\end{equation}

The question that remains is will this model work? To this end, we will use the conductance model for a detailed analysis which can be analogously translated to the resistance model. It turns out for numerical simulations, two issues need be dealt with further, one is computational on ODE solvers and the other is modeling on excitable membrane physiology. Notice from the last three equations of the model that when one or more of the probability functions become zero $\Xphi=0$, it will force in theory the corresponding variable of $n,m,h$ zero as well. But any numerical solver will have a difficulty time to deal with the zero denominator, a so-called stiff solver problem. For this reason we will add a sufficiently small number, $\epsilon>0$, to the denominators inside the square-roots. Otherwise, all numerical solvers we have tried would fail to compute, i.e. converge. This modification solves the stiff-solver problem. As to the modeling problem, notice that if variable $n$, or $m$, or $h$ is zero at sometime $\tau$, any numerical solver applied on the equations will leave it zero for $t\ge \tau$ rather than tracking the limiting probability functions $\Xphi$ which can arise from zero. This is because all standard solvers are build to track only one solution of an initial condition by the uniqueness theorem on differential equations whereas the uniqueness theorem does not apply to our equations. To keep the conductances from being stuck in zero conductance forever because of this inability of all solvers, we will add another sufficiently small number, $\delta>0$, to the numerators inside the square-roots. Coincidentally, this inclusion of the small perturbation can be viewed to model the stochastic phenomenon of spontaneous opening of ion channels. In fact, this small number can be replaced by a small random noise. But the effects are same, keeping the conductances from zero indefinitely. That is, physiologically, the conductances are always above a small value because of the phenomenon of spontaneous firing of ion channels. In conclusion, our final conductance model is as follows:
\begin{equation}\label{eqDengConductanceModel}
\qquad\qquad\qquad\left\{\!\!\!\!\begin{array}{ll} & C\Vvoltageprime=-[\Kbarconductance n (\Vvoltage-\Kabattery)+\Nabarconductance m (\Vvoltage-\Naabattery)+\Gbarconductance h (\Vvoltage-\Gabattery)]\\
& {n}' = \displaystyle \KTimeConstant\sqrt{(n+\delta)/({\Kphi}+\epsilon)}(\Kphi-n)\\
& {m}' = \displaystyle \NaTimeConstant\sqrt{(m+\delta)/{(\Naphi}+\epsilon)}(\Naphi-m)\\
& {h}' = \displaystyle \GTimeConstant\sqrt{(h+\delta)/{(\Gphi}+\epsilon)}(\Gphi-h)\\
\end{array}\right.
\end{equation}
with $\Kphi,\Naphi,\Gphi$ given above. For the resistance model, we only need to cap the resistance functions $\Xpsi$ by a large upper bound to count for the spontaneous opening of ion channels. Although there is no stiffness of the model to deal with, we have to deal with a different kind of solver problem, namely, very large values for variables $x,y,z$ which can slow down the computations because of slower convergence on these variables by any ODE solver.

Figure \ref{figBestModelFit} shows some numerical simulations of the conductance model Eq.(\ref{eqDengConductanceModel}) after it is fitted to some classical experimental data of \cite{hodgkin1952quantitative}. It shows for a set of parameter values, how the solution fits to the experimental data of Hodgkin-Huxley's Axon 17. (The gradient search method used to find this best-fit is the same as described in \cite{deng2017model}.) By comparing to Hodgkin-Huxley equations' fit to the same data as shown in Fig.4(d) and Fig.6(d) of \cite{deng2017model}, one can conclude that our model does no worse. One can even argue that given its mechanistic derivation our model does better than the HH equations. In particular, as shown in Fig.\ref{figBestModelFit}(d), the parallel combination of the sodium and the gating characteristic curves shapes like a letter $N$, automatically giving rise to the negative conductance branch in the middle. How such $N$-nonlinearity arises in neuroscience has always been a puzzling problem (\cite{moore1959excitation,fitzhugh1961impulses}). But for our model it is a simple consequence to the underlining symmetries.

We end this paper by a few remarks. First, there are other functions satisfying the CRKS equation (\ref{eqCRsymmetry}). Specifically, any function of the form $A^i(g)=(\gamma-g)^i/(g\gamma)^{i/2}$ with $i$ being an odd integer is a solution, and any linear combination of two or more of such functions with different $\gamma$ and odd $i$ is also a solution. We only used the simplest form with $i=1$ above. We don't know if this class of functions is the only CRKS solution. Second, is there an ion channel that behaves like a gating channel and therefore can be modeled by CRGS? Such an ion channel is not forbidden by our theory. Third, as shown in Fig.\ref{figBestModelFit}(d), the $IV$-characteristic curve for the potassium channel behaves like a semi-conductor, below $\Kabattery$ it is mostly non=conducting and above $\Kabattery$ it is almost a linear conductor. Notice also that the combined sodium and gating $IV$-characteristic curve behaves like a tunnel diode. Our result may suggest a mathematical model for most nonlinear conductors used in electronics. Fourthly, our model is closely related to a model recently introduced in \cite{deng2017model} (Eq.(8)), whose conductance kinetics can be viewed as an approximation of our model by dropping the square-root factors in the conductance equations for $n, m, h$ in Eq.(\ref{eqDengGRModel}) as we expect them to be near their limiting values $\phi_{ _{\rm K,Na,G}}$ or the square-root factors are near 1. Alternatively, this linear kinetic model can be thought as being derived from the separation of variable condition with $A(g)=\tau (\gamma -g)/g$ whose equivalent form for the resistance kinetics is $dr/dt=\tau r(\rho-r)/\rho$, a nonlinear, logistic equation, implying that it is not kinetically symmetric. This means if a modeler chooses to model the resistance by a linear kinetics $dr/dt=\tau (\rho-r)$, then she will not get the same model if she does the same with the conductance. Likewise, if one chooses to model the resistances by following Hodgkin-Huxley's approach, a different model is sure to arise. Our model Eq.(\ref{eqDengGRModel}) removes this equivocation. This improvement leads to the last point that neurons perhaps can be the consequence to some pure mathematical considerations alone which is quite shocking even if it is only possible. Or evolution of neuron is an unfolding of some elegant symmetries.

\bigskip
\noindent\textbf{Acknowledgement:} The author acknowledges a generous summer visitors fellowship of 2017 from the Mathematics and Science College, Shanghai Normal University, Shanghai, China.

\bibliography{bibcircuit17}
\bibliographystyle{ieeetr}

\end{document}